\documentclass[twocolumn,showpacs,preprintnumbers,amsmath,amssymb,floatfix]{revtex4-1}

\usepackage{graphicx}
\usepackage{amsmath,amsfonts,amssymb}
\usepackage{graphicx}
\usepackage{color}
\usepackage{bbold}

\usepackage[normalem]{ulem}
\usepackage{todonotes}
\usepackage[bookmarks,
            breaklinks,
            pdfstartview=FitH, 
            pdftitle={},
            pdfauthor={},
            pdfkeywords={}
            ]{hyperref}
\usepackage{ulem}
\hypersetup{
  colorlinks   = true,  
  urlcolor     = blue,  
  linkcolor    = blue,  
  citecolor    = red    
}

\usepackage[margin=1.in]{geometry}
\usepackage{graphicx}
\usepackage{amsmath}
\usepackage{amssymb}
\usepackage{mathtools}
\usepackage{amsthm}
\usepackage{tikz-cd}
\usepackage{subfigure}
\usepackage{booktabs}

\begin{document}

\title{The multichannel nature of three-body recombination for ultracold $^{39}$K}
\author{T. Secker}\thanks{These authors contributed equally and share first authorship.}
\affiliation{Eindhoven University of Technology, P.~O.~Box 513, 5600 MB Eindhoven, The Netherlands}
\author{J.-L. Li}\thanks{These authors contributed equally and share first authorship.}
\affiliation{Eindhoven University of Technology, P.~O.~Box 513, 5600 MB Eindhoven, The Netherlands}
\author{P. M. A. Mestrom}
\affiliation{Eindhoven University of Technology, P.~O.~Box 513, 5600 MB Eindhoven, The Netherlands}
\author{S. J. J. M. F. Kokkelmans}
\affiliation{Eindhoven University of Technology, P.~O.~Box 513, 5600 MB Eindhoven, The Netherlands}

\begin{abstract}
We develop a full multichannel spin model in momentum space to investigate three-body recombination of identical
alkali-metal atoms colliding in a magnetic field. 
The model combines the exact three-atom spin structure and realistic pairwise atom-atom interactions.
By neglecting the interaction between two particles when the spectating particle is not in its initial spin state we arrive at an approximate model.
With this approximate model we achieve excellent agreement with the recent precise measurement of the ground Efimov resonance position in potassium-39 close to 33.58 G [Chapurin $et$ $al$., Phys. Rev. Lett. \textbf{123}, 233402 (2019)].
We analyze the limitations of our approximation by comparing to the numerical results for the full system and find that
it breaks down for Feshbach resonances at larger magnetic fields in the same spin channel. 
There the relevant three-body closed channel thresholds are much closer to the open channel threshold, which enhances the corresponding multichannel couplings. Therefore the neglected components of the interaction should be included for those Feshbach resonances.
\end{abstract}

\date{\today}


\maketitle
\section{Introduction}
The Efimov effect describes a three-body scenario with an infinite number of loosely bound trimer states.
Those bound states commonly referred to as Efimov trimers appear
when the two-body interaction is tuned to be resonant and the $s$-wave scattering length $a$ goes through a pole.
The binding energies of these trimers follow a universal scaling relation, 
$E_{n+1}/E_{n} = e^{-2\pi/s_{0}}$ with $s_{0}\approx 1.00624$ for identical bosons \cite{efimov:1970,braaten:2006}. 
This universal scaling law transfers also to other three-body observables such as the values of the scattering length at which the Efimov trimers meet the three-body continuum \cite{braaten:2006} and cause Efimov resonances in the three-body recombination rate.  
Conventionally, the value in scattering length related to the ground Efimov trimer resonance is defined as the three-body parameter $a_{-}$.
It can be experimentally determined by measuring the three-body recombination rate as a function of the scattering length. 
This has been done in a variety of alkali-metal atomic gases \cite{kraemer:2006,pollack:2009,Gross:2009,Gross:2010,zaccanti:2009,Wild:2012,Ferlaino:2011,Berninger:2012} since the pioneering experimental work in Innsbruck \cite{kraemer:2006}. \\
\indent Orginally thought to be a free parameter, most early experimental observations \cite{kraemer:2006,Gross:2009,Gross:2010,Wild:2012,Ferlaino:2011,Berninger:2012} indicated little variation of $a_{-}/r_{\text{vdW}}$  over different Feshbach resonances, spin states and species, with $r_{\text{vdW}}$ \cite{Chin:2010} the van der Waals length of the species considered. Subsequently, a theoretical study \cite{Wang:2012} based on an adiabatic hyperspherical  approach explained that $a_{-}$ is fixed at approximately $-9.7 \, r_{\text{vdW}}$ due to a universal effective three-body potential barrier that arises from pairwise van der Waals interactions. 
These unexpected experimental findings in combination with the successful therotical explanations \cite{Wang:2012,Naidon:2014a,Naidon:2014l}, are referred to as the van der Waals universality of the three-body parameter (TBP).
This universality is believed to break down away from the broad Feshbach resonance limit, as the explanation is based on a single-channel approximation that cannot correctly describe the two-body physics of a narrow Feshbach resonance. 
Previous theoretical investigations based on simplified multichannel models at the two-body level \cite{Petrov:2004,Gogolin:2008,Schmidt:2012} revealed a completely different behavior of $a_{-}$ in the limit of narrow Feshach resonances, indicating the importance of multichannel effects. 
A precise measurement in the $^{39}$K gas in the vicinity of a Feshbach resonance with intermediate width has reported a violation of the single channel van der Waals universality \cite{Chapurin:2019}. 
This finding motivates our current work for developing a full multichannel spin model for three-body recombination. \\
\indent Moreover, recent experimental progress has been made in the field of ultracold chemistry in a regime where the interaction is non-resonant \cite{Harter:2013,Wolf:2017,Wolf:2019}. 
These experiments focus on the distribution of three-body recombination products in ${}^{87}$Rb. 
There the recombination into shallow dimers could be well explained by a single-channel model. 
However, the multichannel structure is required to describe recombination into deeply bound dimer states. \\
\indent  Multichannel effects are inherent in the three-body calculation for a system of ultracold alkali-metal atoms exposed to a magnetic field. 
They are neglected in most calculations due to the intractable complications of solving the three-body equations in that case. Multichannel three-body calculations can be performed by replacing the van der Waals interactions between the atoms with contact or separable interactions \cite{Petrov:2004,Gogolin:2008,Massignan:2008,Naidon:2011,Schmidt:2012,Colussi:2014,Li:2019,Secker:2020}. 
However, such simplifications lead to a low accuracy, which limits the capacity of the corresponding models to analyze multichannel effects especially in view of van der Waals interactions.
So far a number of numerical multichannel models including van der Waals interaction potentials have been developed \cite{Wang:2014,Kato:2017,Chapurin:2019,Xie:2020}. However, these models always limit the spin space of the atoms. \\
\indent In this work, we supplement our three-body multichannel calculation with both the complete realistic spin structure and van der Waals pairwise interactions. 
This allows us to investigate three-body multichannel effects that have not been included in previous studies on a high accuracy level. 
The paper is organized as follows: Section \ref{sec:IIa} introduces the multichannel Hamiltonian of three atoms in a magnetic field. 
Section \ref{sec:IIb} reviews the Alt-Grassberger-Sandhas (AGS) equation and its connection to the three-body recombination rate in the multichannel case. 
Our numerical results for the recombination rate of three $^{39}$K atoms are shown in Section \ref{sec:III}, followed by conclusions in Section \ref{sec:IV}. \\
\section{Theory}
\subsection{The multichannel three-body Hamiltonian}
\label{sec:IIa}
We consider a system of three identical bosonic alkali-metal atoms, where each atom $i$ ($i =1,2,3$) can occupy several internal spin states $|c_i\rangle$ with energies  $E_{c_i}$, accounting for different hyperfine states of the electronic ground configuration \cite{pethick:2008,Chin:2010}. 
These spin states and energies are the eigenstates and eigenvalues of the sum of hyperfine and Zeeman terms, and can be shifted by applying an external magnetic field.
The eigenstates are commonly labeled by the quantum numbers $(f , m_{f}) \equiv c$ that they correspond to at low magnetic field, with $\mathbf{f} = \mathbf{s} + \mathbf{i}$ the sum of the electronic $\mathbf{s}$ and nuclear spin $\mathbf{i}$ \cite{pethick:2008,Chin:2010}.
In the limit of infinite separation the interaction between the atoms vanishes and we get the free Hamiltonian
\begin{equation}
H_0 = \sum_{c_1 c_2 c_3} \left(T + E_{c_1} + E_{c_2} + E_{c_3} \right) | c_1 c_2 c_3 \rangle \langle c_1 c_2 c_3 | \, ,
\end{equation}
with $T$ the kinetic energy operator in the center of mass frame.
However, when the atoms approach each other, the different spin channels $| c_1 c_2 c_3 \rangle $ get coupled by a multichannel interaction potential $V$.
When one atom $k$ is infinitely far separated from the other two atoms $i$ and $j$, the interaction of the pair $(ij)$ can be accurately described by model potentials $V_{ij}$ with a long-range $- C_6 / r_{ij}^6$ van der Waals tail attached to a short-range part, that depends on the nature of the combined electronic spin of atoms $i$ and $j$.
Here $r_{ij}$ is the distance between atoms $i$ and $j$ and we note that we will sometimes use the notation $\alpha = (ij)$ to indicate a certain partition of the three atoms into a pair $(ij)$ and the remaining particle $k$.
When all three atoms $(ijk)$ approach each other, the interaction needs to be adjusted by a genuine short-range three-body potential $V_{ijk}$.
However, it has been demonstrated that $V_{ijk}$ plays only a minor role in low energy three-body collision processes \cite{Wang:2014,Lee:2007} and is therefore often neglected.
We thus take
\begin{equation} 
V \approx \sum_{\substack{i,j=1 \\ i<j}}^{3} V_{ij}  \label{v3} 
\end{equation} 
as an approximation for $V$.
We model the multichannel pairwise interaction as a sum of singlet $V^S_{ij}$ and triplet $V^T_{ij}$ potentials,
\begin{equation} \label{VIJ}
V_{ij}(r_{ij}) = V_{ij}^S (r_{ij}) \mathcal{P}_{ij}^S + V_{ij}^T (r_{ij}) \mathcal{P}_{ij}^T \, ,
\end{equation}
according to the electronic state structure of two alkali-metal atoms. $ \mathcal{P}^S$ and $ \mathcal{P}^T$ denote the projectors on electronic singlet and triplet states, respectively.
For $V^S_{ij}$ or $V^T_{ij}$ we take either Lennard-Jones model potentials with $N$ and $N-1$ $s$-wave bound states, respectively, as done in \cite{Chapurin:2019}, or we take the highly accurate interaction potentials for ${}^{39}$K as presented in \cite{Stephan:2008}. 
Since each constituent $V_{ij}$ conserves $M^{2b}_{ij} = m_{fi} + m_{fj}$, $M^{3b} = m_{f1} + m_{f2} + m_{f3}$ is a good quantum number of the system under the interaction of Eq. (\ref{v3}).  A Full Multichannel Spin (FMS) model in this situation needs to consider all spin channels $\{| c_1 c_2 c_3 \rangle \, | \, M^{3b}=M^{3b}_{\rm{in}} \}$ that have the same $M^{3b}$ as the incoming spin state $M^{3b}_{\rm{in}}$.
Some typical examples of the spin channel energies involved in a three-body collision with $M^{3b} = -3$ for ${}^{39}$K can be found in Fig.~\ref{fig1}(a).
We note that even for a system of three identical bosons even and odd parity dimer channels  couple in the multichannel scenario, which contrasts with the single channel case where such a coupling is absent. This coupling has been observed experimentally \cite{Harter:2013}.
\\
\indent We also analyze a simplified approximate model, which we refer to as the Fixed Spectating Spin (FSS) model.
For that we restrict the pairwise interaction $V_{ij}$ to the incoming spin component $|c^{\rm{in}}_k\rangle$ of the spectating atom and thereby set the interaction to zero in cases where the spin of the spectating atom is not in the incoming component,
\begin{equation} \label{VIJS}
V_{ij}^{\rm{FSS}}(r_{ij}) = V_{ij}(r_{ij}) | c^{\rm{in}}_k\rangle \langle c^{\rm{in}}_k |. 
\end{equation}
In that way only parts of the interaction potential are neglected while the spin structure is kept intact. 
This approximation is valid when the incoming channel is the only open channel and not all three atoms can get in close proximity to each other simultaneously.
In the single-channel scenario it has been demonstrated that a repulsive barrier that arises in the effective three-body potential of the Efimov channel prevents all three atoms being close.
Consequently, the restriction of Eq. (\ref{VIJS}) is a natural choice when the full multichannel three-body problem is too complicated to solve.
Similar restrictions on the spectating atom's spin are implemented in most previous proposals \cite{Jonsell:2004,Wang:2014,Kato:2017,Chapurin:2019}, however in all those models the natural three-body spin structure is also altered or further restricted in course of the approximation.
In the AGS equation below this leads to a transition operator between atoms $i$ and $j$ that projects onto $|c_k^{\rm{in}}\rangle$ and consequently we could as well restrict the complete AGS equation to $c_k = c_k^{\rm{in}}$. 
\subsection{Three-body recombination}
\label{sec:IIb}
Following Ref. \cite{Moerdijk:1995,Lee:2007,Smirne:2007, Secker:2020t}, one can obtain the three-body recombination rate
\begin{align}
K_3(E)
&= \frac{24 \pi m}{ \hbar} (2 \pi \hbar)^6 \nonumber \\
& \phantom{=} \sum_{d,c_d} q_d |{}_\alpha \langle (q_d,c_d), \varphi_d | U_{\alpha 0} (z) | \psi_{\rm{in}} \rangle |^2 
\end{align}
 by evaluating the transition operator element ${}_\alpha \langle (q_d,c_d), \varphi_d | U_{\alpha 0} (z) | \psi_{\rm{in}} \rangle$, from a free incoming state $\psi_{\rm{in}}$ of energy $E$ into a $\alpha$-dimer $d$ with wave function $\varphi_d$ and energy $E_d$ plus a free atom of spin $c_d$ and absolute momentum $q_d$ relative to the dimer center-of-mass, on the energy shell.  
This leads to $E = 3 q_d^2 / 4 m + E_{c_d} + E_d $ and the complex energy $z = E + \text{i}0$, which means that we take the limit in $z$ from the upper half of the complex energy plane.
Here $m$ denotes the mass of a single atom.  

The transition operator $U_{\alpha 0}(z)$  related to three-body recombination into the $\alpha$-dimer state is defined by the AGS equation for three identical bosons \cite{Lee:2007,Mestrom:2019,Secker:2020,Secker:2020t}
\begin{align} \label{eq:AGSeqrecom}
U_{\alpha 0} (z) & = \frac{1}{3}G_0^{-1}(z) \left[1 + P_+ + P_-\right] \nonumber \\
& \phantom{=} + \left[ P_+ + P_- \right] \mathcal{T}_\alpha (z) G_0(z) U_{\alpha 0} (z) \, .
\end{align}
The operators $G_0$ and $\mathcal{T}_\alpha$ denote the free Greens operator and a generalized two-body transition operator, respectively.
The operators $P_+$ and $P_-$ are the cyclic and anticylclic permutation operators, respectively, that act on both coordinates and spins, since
in a situation with realistic spin structure which we are considering in this work all operators in the AGS equation need to be generalized to account for the multichannel spin structure.
The three-body free Greens operator $G_0$ includes the shifts in the thresholds in different three-body spin channels
\begin{eqnarray}
G_0 (z) &=& (z - H_0)^{-1} \nonumber \\
&=&\sum_{c_1,c_2,c_3}\frac{| c_1 c_2 c_3 \rangle \langle c_1 c_2 c_3 |}{z-E_{c_1} -E_{c_2} - E_{c_3}-T}. \label{G0}
\end{eqnarray}
From Equation (\ref{G0}) we can infer, that $G_0$ can serve as a suppressing factor in channels with large threshold difference to the incoming channel $\Delta E = \sum_{i=1}^3 ( E_{c_i} -E_{c_i^{\rm{in}}}) \gg E_{\mathrm{vdW}}$ in the low energy and low momentum regime where both $z - \sum_{i=1}^3E_{c_i^{\rm{in}}}$ and $T$ are small.
Here $E_{\mathrm{vdW}} = \hbar^2 / m r_\mathrm{vdW}^2$ is the van der Waals energy.
We interpret our numerical three-body results close to Feshbach resonances with varying threshold differences in view of the suppressing properties of $\Delta E$ in section \ref{sec:III}.
The generalized two-body transition operator $\mathcal{T}_\alpha $ is given by
\begin{eqnarray}
\mathcal{T}_\alpha (z) &=& (1 - V_\alpha G_0 (z))^{-1} V_\alpha  \\
&=&\sum_{c_k}\int d\mathbf{q}t(z-E_{c_k}-\frac{3q^2}{4m})|c_k,\mathbf{q}\rangle_{\alpha \alpha}\langle c_k,\mathbf{q}|, \nonumber
\end{eqnarray}
where $t(z-E_{c_k}-3q^2/4m)$ is the two-body transition operator acting on the spin and relative coordinate of the pair of atoms $(ij)$. The relative momentum between the center of mass of the pair $(ij)$ and the atom $k$ is denoted by $\mathbf{q}$.
The partial wave components of $t(z-E_{c_k} - 3q^2/4m)$ can be obtained by extending the method in Ref. \cite{Secker:2020t} to the multichannel case. 
In the following we will omit the explicit dependence on $z$ for notational compactness unless it is needed.
 
In the on-shell limit $z = E + \mathrm{i} 0$, it is more convenient to define a new operator $A_\alpha$ \cite{Secker:2020t}
 \begin{equation}
A_\alpha = 3 G_0 \left( P_+ + P_- \right) \mathcal{T}_\alpha  G_0  U_{\alpha 0} \, ,
\end{equation}
which fulfills the following equation
 \begin{equation} \label{eq:AGSAalpha}
A_\alpha = G_0 \left( P_+ + P_- \right) \mathcal{T}_\alpha \left[ 1 + P_+ + P_- + A_\alpha \right]\, .
\end{equation}
as a consequence of Eq. (\ref{eq:AGSeqrecom}).
Since the inhomogeneous term in $U_{\alpha0}$ evaluates to zero when applied on $|\psi_{\rm{in}}\rangle$, we get
\begin{equation}
{}_\alpha \langle (q_d,c_d), \varphi_d | U_{\alpha 0} (z)|  \psi_{\rm{in}} \rangle=\frac{1}{3}{}_\alpha \langle (q_d,c_d), \varphi_d |V_{\alpha}A_\alpha| \psi_{\rm{in}} \rangle 
\end{equation}
in the on-shell limit (see Ref. \cite{Secker:2020t} for more details), so that we can consider Eq. (\ref{eq:AGSAalpha}) instead of Eq. (\ref{eq:AGSeqrecom}) to obtain $K_3$.
 
We expand $\mathcal{T}_\alpha = \int dqq^{2}\sum_i \tau_\alpha (i,q) | i, q \rangle_{\alpha} \langle i,q |$ and use the incoming state $|\psi_{\rm{in}}\rangle$, such that we arrive at the linear system
\begin{align} \label{eqa}
& {}_{\alpha}\langle q', i' | A_\alpha | \psi_{\rm{in}} \rangle  \\
& = \int dqq^{2}\sum_i \langle q',i' | G_0 (P_+ + P_-) | q,i\rangle_{\alpha} \tau_\alpha (q,i)  \nonumber \\
& \phantom{=} \left[ {}_{\alpha}\langle q,i| (1 + P_+ + P_-) | \psi_{\rm{in}} \rangle +  {}_{\alpha}\langle q,i| A_\alpha | \psi_{\rm{in}} \rangle \right] \, . \nonumber
\end{align}
The partial three-body recombination rates are then obtained by evaluating the on-shell elements of ${}_{\alpha}\langle q, i | A_\alpha | \psi_{\rm{in}} \rangle$, since the expansion base ${}_{\alpha}\langle q, i |$ naturally includes the terms ${}_\alpha \langle (q_d,c_d), \varphi_d |V_\alpha$. 
It should be noted that the multichannel structure of Eq. (\ref{eqa}) is implicitly contained in all operators and state vectors. 
Therefore Eq. (\ref{eqa}) is a multichannel generalization of the corresponding equation in Ref. \cite{Secker:2020t} even though both look identical.
More details on the linear system can be found in appendix \ref{A}. \\

\section{Results}
\label{sec:III}
   \begin{table*}
 \tabcolsep=10pt
\small
\renewcommand\arraystretch{1.0}
\caption{\label{tab:1}Comparison of three-body parameters from Ref. \cite{Chapurin:2019} and our FSS calculation (this work) with Lennard-Jones (LJ) potentials supporting different number $N$ of singlet $s$-wave dimer states. Our results with realistic $^{39}$K molecular potentials (full)  and the experimental measurement \cite{Chapurin:2019} are also shown.  A momentum cutoff at $q_{\rm{max}}$ = 20 $\hbar/r_{\rm{vdW}}$ is implemented in all calculations in this work except for the ‘full40’ case, where $q_{\rm{max}}$ = 40 $\hbar/r_{\rm{vdW}}$ instead. For definiteness, the singlet and triplet scattering lengths specifying the potentials we use are listed in the last two columns. }
 \begin{tabular}{ccccccc}
 \hline
 \hline
 &\multicolumn{2} {c}{$a_{-}$[$r_{\text{vdW}}$]}&\multicolumn{2} {c}{$\eta$}\\
 &this work&Ref. \cite{Chapurin:2019}& this work&Ref. \cite{Chapurin:2019} & $a^{S}$ [$r_{\text{vdW}}$] &$a^{T}$ [$r_{\text{vdW}}$]\\
 \hline
 $N$=2, LJ&$-13.51$&$-7.61$& $0.23$&$0.10$&$1.3503$&$-0.5098$  \\
 $N$=3, LJ& $-14.33$&$-11.20$ &$0.20$&$0.19$&$1.7281$&$-0.5238$ \\
  $N$=4, LJ& $-14.16$&$-12.27$ & $0.18$&$0.20$&$1.8479$&$-0.5238$ \\
 $N$=5, LJ&$-14.60$&$-12.69$ &$0.20$&$0.21$&$1.9036$&$-0.5238$\\
 full &$-14.12$&$-$&$0.15$&$-$ &$2.1432$&$-0.5181$\\
 full40&$-14.03$&$-$&$0.19$&$-$ &$2.1432$&$-0.5181$ \\
measurement&\multicolumn{2} {c}{-14.05 (17)}&\multicolumn{2} {c}{0.25 (1)} \\
 \hline
 \hline
 \end{tabular}
 \end{table*}

\begin{table*}
 \tabcolsep=10pt
\small
\renewcommand\arraystretch{1.0}
 \caption{\label{tab:2} Comparison of three-body parameters from the FSS and FMS calculations with $N=2$ Lennard-Jones potentials concentrating on the same Feshabach resonance as in Table \ref{tab:1}. $\ell_{\rm{max}}$ denotes the maximum dimer partial-wave included in the calculation. The FSS result with realistic molecular potential and $q_{\rm{max}}=40$/$r_{\rm{vdW}}$ (full40) is also shown. }
 \begin{tabular}{ccccccc}
 \hline
 \hline
&\multicolumn{2}{c}{FSS ($N=2$, LJ)}& \multicolumn{2}{c}{FMS ($N=2$, LJ)}& \multicolumn{2}{c}{FSS (full40)}\\
$\ell_{\text{max}}$&$a_{-} [r_{\text{vdW}}]$&$\eta$&$a_{-}[r_{\text{vdW}}]$&$\eta$&$a_{-}[r_{\text{vdW}}]$&$\eta$  \\
\hline
$0$&$-14.19$&$0.14$&$-18.34$&$0.17$&$-14.52$&$0.16$\\
$2$&$-15.73$&$0.17$&$-20.66$&$0.06$&$-15.34$&$0.15$\\
$4$ &$-13.90$&$0.20$&$-14.32$&$0.15$&$-14.89$&$0.19$\\
$6$&$-13.60$&$0.23$&$-13.50$&$0.20$&$-14.42$&$0.18$\\
$8$&$-13.55$&$0.23$&$-13.37$&$0.19$&$-14.25$&$0.19$\\
$10$&$-13.51$&$0.23$&$-$&$-$&$-14.03$&$0.19$ \\
 \hline
 \hline
 \end{tabular}
 \end{table*} 
Recently, three-body recombination rates were precisely measured in a $^{39}$K atomic gas near a Feshbach resonance at  33.58 G. 
The three-body parameters $a_{-}$ and $\eta$, characterizing the position and width of the lowest Efimov resonance, are extracted and confirmed by a state-of-the-art adiabatic hyperspherical calculation \cite{Chapurin:2019}. 
The measurement was found to be in agreement with a previous experimental result \cite{Roy:2013}.
This poses ideal circumstances for checking our new numerical method. 
For that we have to consider a system of three $^{39}$K atoms initially prepared in the  $|f_{1}=1,m_{f 1}=-1\rangle|f_{2}=1,m_{f 2}=-1\rangle|f_{3}=1,m_{f 3}=-1\rangle$ state in an external magnetic field $B$.

We first use the FSS model to obtain the position $a_-$ and width $\eta$ of the lowest Efimov resonance of  three $^{39}$K atoms nearby $B= 33.58$ G. 
For that, we calculate the three-body recombination rate $K_3(0)$ by solving Eq.~(\ref{eqa}) at zero energy for a set of magnetic fields close to the lowest Efimov resonance and fit the universal expressions for $K_3(0)$ \cite{braaten:2006} to our results. 
In our numerical computations, we restrict to a maximum of $\ell_\mathrm{max} = 10$ partial waves in the atom-dimer momentum and to an integration range in $q$ of $[0,q_\mathrm{max}]$.  We implement $q_\mathrm{max} = 20$ $\hbar/ r_\mathrm{vdW}$ in our calculations if not specified differently. We optimize the parameters of our Lennard-Jones potential model to best represent the two-body scattering length $a$ in the $M^{2b}_{ij}=  -2$ channel close to the Feshbach profile with magnetic field points being sampled in  both background and resonance regimes. 
Table \ref{tab:1} shows our results for Lennard-Jones potentials with $N$ singlet $s$-wave dimer states and for the realisitic molecular potentials including Born-Oppenheimer corrections from Ref. \cite{Stephan:2008}.
For the Lennard-Jones potential our results show just little variation of $a_-$ when the number of $s$-wave dimers supported by the singlet potential is changed from $N=2$ to $N=5$, in contrast to the constant decrease of $a_-$ that has been found for the model in Ref. \cite{Chapurin:2019}. Qualitatively, the behavior we find is more similar to that of a single channel model \cite{Wang:2012}. 
When using the realistic molecular potentials, we find $a_- \approx -14$ $r_{\rm{vdW}}$, in excellent agreement with the experimental result  $a_-=-14.05 (17)$ $r_{\rm{vdW}}$.

We find that $\ell_\mathrm{max} = 10$ is sufficient for a good convergence in the FSS model, as can be seen from the $\ell_{\rm{max}}$-dependence of the FSS model results presented in Table \ref{tab:2}. 
The error resulting from the restriction in integration range has been analyzed for single channel Lennard-Jones potentials with close to 4 and 6 bound $s$-wave dimer states \cite{Secker:2020t}. 
There the error in $a_-$ is demonstrated to be 2 $\%$ for $q_\mathrm{max} = 20$ $\hbar/ r_\mathrm{vdW}$ and 0.5 $\%$ for $q_\mathrm{max} = 40$ $\hbar/ r_\mathrm{vdW}$.
We estimate that the error from restricting the integration range is of similar order of magnitude in our present calculations. 
In this view, our results for $a_-$ are reliable up to uncertainties of a few percent. The error in $\eta$, however, could be large according to our analysis in \cite{Secker:2020t}.  

\begin{table*}
 \tabcolsep=10pt
\small
\renewcommand\arraystretch{1.0}
 \caption{\label{tab:3} Three-body parameters from FSS and FMS calculation with $N=2$ Lennard-Jones or realistic $^{39}$K molecular potential (full). The upper and lower panels list the results for FR2 at  162.35 G and FR3 at 560.72  G, repectively.}
 \begin{tabular}{ccccccc}
 \hline
 \hline
& \multicolumn{2}{c}{FSS ($N=2$, LJ)}&\multicolumn{2}{c}{FMS ($N=2$, LJ)}&\multicolumn{2}{c}{FMS (full)}\\
$\ell_{\text{max}}$&$a_{-} [r_{\text{vdW}}]$&$\eta$&$a_{-}[r_{\text{vdW}}]$&$\eta$&$a_{-}[r_{\text{vdW}}]$&$\eta$ \\
\hline
$0$&$-31.22$&$0.12$&$-24.34$&$0.11$&$-29.74$&$0.16$\\
$2$&$-32.68$&$0.40$&$-9.83$&$0.29$&$-27.29$&$0.20$ \\
$4$&$-35.70$&$0.57$&$-38.48$&$0.29$&$-25.05$&$0.20$\\
$6$&$-33.85$&$0.53$&$-29.23$&$0.33$&$-24.33$&$0.21$\\
$8$&$-33.79$&$0.54$&$-27.53$&$0.33$&$-24.14$&$0.22$\\
$10$&$-33.81$&$0.54$& $-$&$-$&$-$&$-$\\
\hline
$0$&$-18.91$&$0.16$&$-19.07$&$0.06$&$-27.17$&$0.14$\\
$2$&$-16.73$&$0.02$&$-11.30$&$0.26$&$-19.32$&$0.13$ \\
$4$&$-15.42$&$0.03$&$-24.14$&$0.71$&$-15.93$&$0.06$ \\
$6$&$-15.00$&$0.04$&$-19.28$&$0.47$&$-14.76$&$0.09$\\
$8$&$-14.85$&$0.04$&$-18.54$&$0.45$&$-13.98$&$0.11$\\
$10$&$-14.78$&$0.04$&$-$&$-$&$-$&$-$ \\
 \hline
 \hline
 \end{tabular}
 \end{table*}
 \begin{figure*}[htb!] 
	  \centering
	  \includegraphics[width = 1.0 \textwidth]{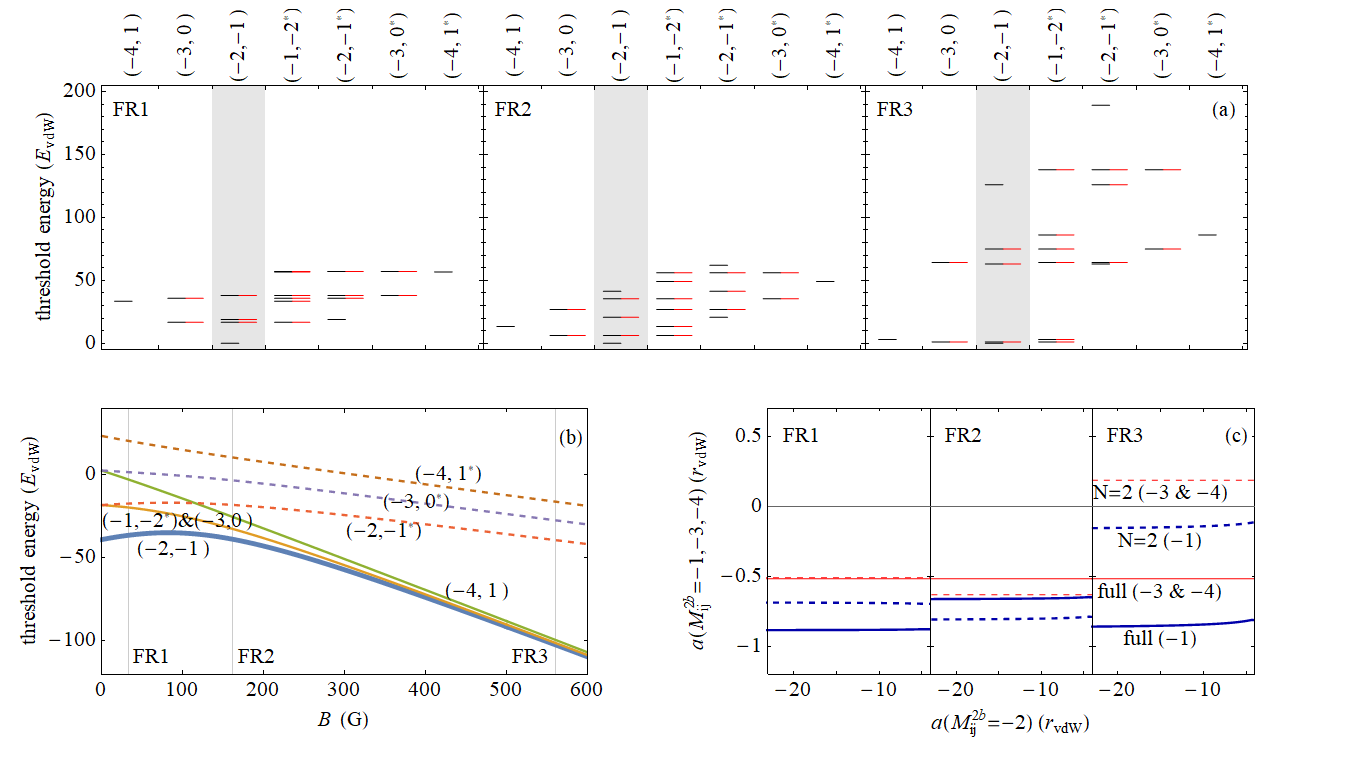}
\caption{\label{fig:1} (a) Threshold energies relative to the incoming channel of all spin channels involved in the FMS model for the different Feshbach resonances analyzed. 
The spin channels are grouped by the quantum numbers ($M_{ij}^{2b}, m_{f_k}$) along the $x$-direction.
The $m_{f_k}$ and  $m_{f_k}^*$ correspond to the $|f_k=1,m_{f_k}\rangle$ and $|f_k=2,m_{f_k}\rangle$ states, respectively and we distinguish thresholds related to even and odd parity by black and red coloring, respectively. 
We note that threshold lines with the same $x$-position are directly coupled by $V_{\alpha}$.
The gray shaded area indicates the thresholds that are included in the FSS model.
(b) Lowest threshold energies with ($M_{ij}^{2b}, m_{f_k}$)  involved in the colliding system of three $^{39}$K atoms with $M^{3b} = -3$.
The vertical gray lines denote the Feshbach Resonance at 33.58 G (FR1), 162.35 G (FR2) and 560.72 G (FR3). (c) Two-body scattering length with different $M_{ij}^{2b}$ as a function of the incoming channel two-body scattering length $a (M_{ij}^{2b} =-2)$ from the fitted ($N=2$) Lennard-Jones potentials (dashed) and realistic molecular potentials (full). 
For three-body calculations with $M^{3b} = -3$ the two-atom spin states involved have $M_{ij}^{2b} = -1, -3, -4$.
}
\label{fig1}
\end{figure*}

To go beyond the FSS approximation, we consider the FMS model of the three atom system.
For the $N=2$ Lennard-Jones potentials representing the Feshbach resonances at 33.58 G in the $M^{2b}_{ij} = -2$ channel,  we perform the calculation in the FMS model  and find $a_-$ in good agreement with the FSS results (see Table \ref{tab:2}), indicating that the FSS model is a good approximation when considering $a_-$. 

However, considering the Feshbach resonances at larger magnetic fields of 162.35 and 560.72 G in the $M^{2b}_{ij} = -2$ channel represented by $N = 2$ Lennard-Jones potentials we find that the FSS approximation is breaking down. 
That is apparent by comparing the FSS and FMS results in Table \ref{tab:3}. 
A possible reason for this could be that the spin states neglected in the FSS model are less suppressed in these cases.
A suppression in the coupling to those spin channels arises due to the separation in channel energy, which we commented on after Eq. (\ref{G0}).
As is depicted in Figs. \ref{fig:1}(a) and \ref{fig:1}(b) this separation is much larger for the resonance at  33.58 G (FR1) than for the one at 162.35 (FR2) and 560.72 G (FR3). 
Therefore we suspect that the FSS model works for the FR1 but the FMS model has to be used for FR2 and FR3. 

Even though our $N=2$ Lennard-Jones FMS results are not completely converged for FR2 and FR3, they tend to be in disagreement with the experimental observations of $a_-= -11.3 (1.9)$ $r_{\rm{vdW}}$ for FR2 and $a_-=-9.9 (1.4)$ $r_{\rm{vdW}}$ for FR3 \cite{Roy:2013}.
A reason could be that the pairwise interactions in the two-body channels with $M_{ij}^{2b} = -1, -3$ and $-4$ are not accurately represented.
Since the Lennard-Jones potentials are just adjusted to represent the interactions in the $M^{2b}_{ij} = - 2$ channel correctly, it is not surprising that the interactions in the other two-body channels deviate from the original interactions. 
This can be seen when comparing the corresponding two-body scattering lengths shown in Fig.~\ref{fig:1}(c), in which we find deviations for all three cases. 
The deviations in case of FR1 are not causing problems since the contribution from the inaccurately represented interactions can be neglected.

We also give the FMS results for the realistic molecular potentials in Table \ref{tab:3}, even though they are not converged in the partial waves included. 
The results including up to $8$ partial waves already indicate a better agreement with the experimental results for the resonances at 162.75 and 560.72 G than the Lennard-Jones potential model.
We note that for the realistic potentials the interactions in all spin states are correctly represented. 

\section{conclusion and outlook}
\label{sec:IV}
In summary, we study multichannel effects on three-body recombination of ultracold $^{39}$K atoms. We solve the three-body equation with van der Waals pairwise interactions including the realistic spin structure of the system. 
We numerically confirm that restricting the pairwise interaction to the incoming spin component of the spectating atom, i.e. the FSS model, is a good appoximation around the Feshbach resonance at 33.58 G in the $M^{2b}_{ij} = -2$ channel. 
The FSS model gives $a_- \approx -14$ $r_{\rm{vdW}}$ in excellent agreement with the current precise measurement $a_- = -14.05 (17)$ $r_{\rm{vdW}}$ \cite{Chapurin:2019}. 
In addition, we analyzed the limitations of the FSS model by investigating two other Feshbach resonances at 162.35 and 560.72 G.
There the three-body channels neglected in the FSS model are less suppressed due to the smaller threshold difference to the open channel and need to be taken into account.
In such a situation, the interactions in all contributing channels should be well represented to arrive at an accurate full multichannel spin model.

The results of this work raise doubts on the approximation of restricting or altering the realistic three-atom spin structure, which has been implemented earlier to enable multichannel three-body numerical calculations \cite{Jonsell:2004,Wang:2014,Kato:2017,Chapurin:2019}.
To determine in which situations this approximation is valid and when it is not is an important task, which needs to be further clarified in future investigations.
Nonetheless, we gained some preliminary insight. At large threshold differences, three-body multichannel couplings to the neglected channels can be suppressed. This could be one indicator for the regime in which the approximation holds. 
In this view, it can be expected that heavier species, such as Rb and Cs for which the threshold differences relative to the van der Waals energy scale are generally larger, are better represented by the approximation than lighter ones, such as Na and Li. 
This may explain that agreement has been achieved with multichannel models for Rb and Cs systems \cite{Wang:2014}, but not for the Li system.  
In addition, the strength of the two-body multichannel coupling which is related to the width parameter $s_{\rm{res}}$ \cite{Chin:2010} of a Feshbach resonance, may also affect the validity of the approximation.
A more rigorous  treatment should take both the threshold difference and $s_{\rm{res}}$ into account. 

\section*{Acknowledgements}
We thank Jos\'e D'Incao, Denise Ahmed-Braun, Victor Colussi, Gijs Groeneveld, and Silvia Musolino for discussions.
This research is financially supported by the
Netherlands Organisation for Scientific Research (NWO)
under Grant No. 680-47-623.

\appendix

\section{Details for solving Eq. (\ref{eqa})} \label{A}
In order to solve  Eq. (\ref{eqa}), one needs to evaluate $\langle q',i' | G_0 (P_+ + P_-) | q,i\rangle_{\alpha}$ and ${}_{\alpha}\langle q,i| (1 + P_+ + P_-) | \psi_{\rm{in}} \rangle$, which requests $| q,i\rangle_{\alpha}$ at first. We note that $| q,i\rangle_{\alpha}$ is a multichannel analogue as that in Ref. \cite{Secker:2020t} and can be calculated by the same mapped grid Hamiltonian approach \cite{Willner:2004}. To be explicit,
\begin{equation} \label{i}
 |i\rangle=\sum_{C}|\chi(p;C,n,q,c_k,\Lambda)\rangle| C\rangle| c_k\rangle|\Lambda\rangle, 
 \end{equation}
 where $|\Lambda\rangle=|L \ell J M_{J}\rangle$ denote quantum numbers for all spatial angular momenta, say, $L$ for the atom-dimer orbit, $\ell$ for the dimer orbit, $J$ and $M_{J}$ for the total orbit. $|C\rangle$ describes the spin state $|c_ic_j\rangle$ of pair ($i,j$), which is chosen to be symmetric for even $\ell$ and antisymmetric for odd $\ell$ considering the identical bosonic system in this work. Note that the summation over $C$ is under the restriction of $M_{ij}^{2b}+m_{f_k}= M^{3b}$. $|\chi(p;C,n,q,c_k,\Lambda)\rangle$ is the $C$ channel component of $n$th eigenstate of two-body $\ell$th partial-wave transition operator, which is obtained at $E^{\text{2b}}=E-E_{c_{k}}-3q^{2}/4m$. In terms of the two-body spin basis $|C\rangle$, it can be proved that $P_+$ and $P_-$ contribute equally in Eq. (\ref{eqa}) as is done in Ref. \cite{Glockle:1983}.  In the following, we shall replace $P_+ + P_-$ by 2$P_+$ in $\langle q',i' | G_0 (P_+ + P_-) | q,i\rangle_{\alpha}$ and ${}_{\alpha}\langle q,i| (1 + P_+ + P_-) | \psi_{\rm{in}} \rangle$.
 
 \indent The zero energy incoming state can be written as 
 \begin{equation}
 |\psi_{\rm{in}} \rangle=|q=0,p=0,C^{\rm{in}},c_{k}^{\rm{in}},\Lambda=0\rangle,
 \end{equation}
 where $\Lambda = 0$ means $L=0, \ell=0, J=0$ and $M_J=0$.
 So that ${}_{\alpha}\langle q,i| (1 + P_+ + P_- ) | \psi_{\rm{in}} \rangle$ is given by
 \begin{eqnarray}
&&{}_{\alpha}\langle q,i| (1 + 2P_+ ) | \psi_{\rm{in}} \rangle = \sum_{C}\chi(0;C,n,0,c_k^{\rm{in}},0) \nonumber \\ 
&& \times\frac{\delta(q)}{q^2}\langle Cc_k|(1+2P_+^{\rm{s}})|C^{\rm{in}}c_k^{\rm{in}}\rangle\delta_{\Lambda0},
 \end{eqnarray}
 where $P_+^{\rm{s}}$ is $P_+$ acting on spin space. $\langle q',i' | G_0 (P_+ + P_-) | q,i\rangle_{\alpha}$ is more complicated, which consists of a lot inner products of  $\langle\chi',C',n',q', c'_k,\Lambda' |G_0P_+|\chi,C,n,q,c_k,\Lambda\rangle$ according to Eq. (\ref{i}). Due to the conservation of total angular momentum of three atoms, we have $|\Lambda\rangle=|\ell,\ell,0,0\rangle$ and then get
\begin{widetext}
\begin{eqnarray}
&&\langle\chi',C',n',q', c'_k,\Lambda' |G_0P_+|\chi,C,n,q,c_k,\Lambda\rangle \nonumber \\
&&=\langle\chi',n',q',\Lambda'|\hat{G}_{0}(E-E_{C}-E_{c_k})\hat{P}_{+}^{\text{c}}|\chi ,n,q,\Lambda\rangle\langle C'c'_k|(P_+^{\rm{s}})|Cc_k\rangle  \nonumber \\
&&=\frac{(-1)^{\ell}\sqrt{2\ell'+1}\sqrt{2\ell+1}}{2}\int_{-1}^{1}du P_{\ell'}\left(\frac{q'^{2}/2+q'qu}{q'\sqrt{q'^{2}/4+q^{2}+q'qu}}\right) P_{\ell}\left(\frac{q^{2}/2+q'qu}{q\sqrt{q^{2}/4+q'^{2}+q'qu}}\right)\nonumber \\ 
&&\times \frac{\left[\chi(\sqrt{q'^{2}/4+q^{2}+q'qu};C',q',n',c'_k,\Lambda')\right]^{*}\chi(\sqrt{q^{2}/4+q'^{2}+q'qu},C,n,q,c_k,\Lambda)}{E+\text{i}0-E_{C}-E_{c_{k}}-q'^{2}/m-q^{2}/m-q'qu/m} \langle C'c'_k|(P_+^{\rm{s}})|Cc_k\rangle, \label{ZZ}
\end{eqnarray}
\end{widetext}
where $P_+^{\rm{c}}$ is $P_+$ acting on coordinate space and $P_\ell$ the Legendre polynomial.

\bibliography{biblio-TS}

\end{document}